\documentclass[doublecol]{epl2} 

\usepackage{graphicx}
\usepackage{amsmath}
\usepackage{amssymb}
\usepackage{units}

\title{Spin-boson dynamics: A unified approach from weak to strong coupling}

\author{Francesco Nesi\inst{1} \and Elisabetta Paladino\inst{2} \and Michael Thorwart\inst{3} \and Milena Grifoni\inst{1}}

\institute{                    
  \inst{1} Theoretische Physik, Universit\"at Regensburg, 93040 Regensburg,
Germany\\
  \inst{2} MATIS INFM-CNR \& Dipartimento di Metodologie Fisiche e Chimiche,
Universit\`a di Catania, 95125 Catania, Italy\\
  \inst{3} Institut f\"ur Theoretische
Physik, Heinrich-Heine-Universit\"at D\"usseldorf, 40225  D\"usseldorf, Germany
}
\pacs{05.30.-d}{Quantum statistical mechanics}
\pacs{03.65.Yz}{Decoherence; open systems; quantum statistical methods}
\date{\today}

\abstract{
We present a novel approximation scheme to describe the influence of a harmonic bath on the dynamics of a two-level particle over almost the whole regime of temperatures and coupling to the environment, for a wide class of bath spectral densities. Starting from the exact path-integral solution~for the two-level system density matrix, effective intra-blip correlations are fully included, while inter-blip and blip-sojourn interactions are considered up to first order. In the proper regimes, an excellent agreement with conventional perturbative approaches and ab-initio path-integral results is found.}
\begin{document}
\maketitle
\section{Introduction}

The problem of a two-level system (TLS) suffering from
environmental decohering effects is ubiquitous to many physical
and chemical situations \cite{Leggett87,Weiss99,PhysRep98}.
Standard examples involve electron and proton transfer reaction in
condensed phases \cite{Garg85}, defect tunneling in metals \cite{Golding} or tunneling systems in glasses \cite{glass, glass2}. 
% , or two-level atoms in an optical cavity \cite{optical}. 
Recently, several
realizations of TLSs have been experimentally demonstrated in superconducting
  \cite{Nakamura99} and semiconducting \cite{Hayashi03} devices as possible unit (quantum bit)
 for future quantum computers. In these solid state systems, decoherence is a major obstacle towards the realization of a
 usable quantum computer \cite{Makhlin01,Vorojtsov05,Goorden04}.
Hence, a proper understanding of dissipation over a broad parameter regime is of outermost importance.

 For a description of the dissipative dynamics the
spin-boson model, in which the TLS is bilinearly coupled to a harmonic bath, is very frequently used. 
It reads  \cite{Leggett87,Weiss99,PhysRep98}
\vspace{-0.05cm}\begin{equation}
\hat
H(t)=\frac{\hbar}{2}[\varepsilon(t)\hat\sigma_z-\Delta\hat\sigma_x]-\frac{1}{2}\hat\sigma_z\hat
X+\hat H_{\rm B }\;. \label{spin-boson}
\vspace{-0.05cm}\end{equation}
 %When having the persistent current qubit in mind \cite{Mooij99},
The basis states
 $|{\rm R\rangle}$ and $|{\rm L}\rangle$
 are the localized eigenstates of the ''position'' operator
$\hat\sigma_z$, $\Delta$ describes the coupling between the
two-states due to tunneling, and $\varepsilon (t)$ is an external
control field. The Hamiltonian $\hat H_{\rm
B}=\sum_i\hbar\omega_i(\hat b^\dagger_i \hat b_i+1/2)$ represents
a bath of bosons, and the collective variable $\hat X=\sum_i
c_i(\hat b_i+\hat b_i^\dagger)/2$ describes the bath polarization.
Despite
the huge amounts of works on the subject
\cite{Leggett87,Weiss99,PhysRep98}, the existing schemes for a
portrayal of the time-evolution of the TLS reduced density matrix mostly reduce to two main roads of approximation.
 On the one hand the so termed noninteracting-blip approximation (NIBA) \cite{Leggett87,Weiss99},
 or equivalent projection operator techniques \cite{Aslangul86}
  based on an expansion to leading order in the tunneling matrix element $\Delta$,
  has been proved to be  successful
  in the regimes of high temperatures and/or strong friction.
On the other hand the weak coupling and low-temperature regime,
where NIBA fails for an asymmetric TLS, is typically tackled
within an expansion to lowest order in the TLS-bath coupling. In
this latter case path-integral methods \cite{EPJB99,Görlich89} as well as
the Bloch-Redfield formalism are used \cite{Argyres64}
 (the two
methods have been demonstrated to  yield the same dynamics for
weak Ohmic damping \cite{Hartmann00}), or a Born approximation \cite{Loss05}. To date, only numerical ab-initio
calculations \cite{Egger,Makri95,Stockburger,Goorden04} can provide
a description of the TLS dynamics smoothly interpolating between a
weak and a strong coupling situation.

In this work, we present an interpolating approximation scheme, enabling to describe the weak and strong coupling regimes
in a unique scheme. We call it weakly-interacting blip approximation
(WIBA),
within which the dynamics of the population difference~$\langle \hat{\sigma}_z \rangle_t \equiv P(t)$ is%=P_R(t)-P_L(t)$ reads
\vspace{-0.15cm}\begin{equation}
%\begin{split}
\dot P(t)=-\int_{0}^t dt'[
K^a(t,t')-W(t,t') + K^s(t,t')P(t')]\;.
% \sigma_x(t)&=
% \int_{0}^tdt'[K^s_x(t,t')+K_x^a(t,t')\sigma_z(t')]\;, 
\label{GME}
\vspace{-0.15cm}%\end{split}
\end{equation}
The irreducible kernels $K^{(s/a)}$, $W$ %$K_x^{(s/a)}$,
entering this generalized master equation %(GME)
 are
 neither perturbative  in the tunneling matrix  $\Delta$ nor in the TLS-bath coupling, and are given in analytical form in \eqref{WIBAs}, \eqref{WIBAa} and \eqref{WIBAw} below. %In particular, $W$ depends on the initial preparation of the system.
By comparing the predictions of the WIBA with known perturbative results as well as with exact ab-initio calculations, we show that the WIBA well describes the TLS dynamics over the whole regime of temperature and environmental coupling.

In the spin-boson model all the effects of the bath on the TLS are captured
by the spectral density $G(\omega)=\pi\hbar^{-2}\sum_i c_i^2\delta
(\omega-\omega_i)$. In the following we shall consider a class of
spectral densities with a continuous  spectrum:%, such that~\cite{Weiss99}
\vspace{-0.05cm}\begin{equation}
 G(\omega)=2\delta_s \omega_{\rm{ph}}^{1-s}\omega^s e^{-|\omega|/\omega_c} \;, \label{deltas}\vspace{-0.05cm}
\end{equation}
with $\delta_s$ being a dimensionless coupling parameter,
$\omega_{\rm{ph}}$ a characteristic phonon frequency, and $\omega_c$
the bath cut-off frequency. Thus, \eqref{deltas} encompasses the
commonly considered Ohmic spectrum ($s=1$) \cite{Leggett87,Weiss99,PhysRep98,Garg85,Golding,Aslangul86,Hartmann00,Loss05,Egger, Stockburger,Annals}, with
$\delta_1=\alpha$ being the so-called TLS Kondo parameter, and the super-Ohmic case \cite{Leggett87,Weiss99,glass,glass2,Vorojtsov05}.
The applicability of the WIBA to other classes, as e.g. structured baths \cite {Garg85,Goorden04}, will be discussed elsewhere.

\section{Exact path-integral formulation}

To start with, we assume a factorized initial
condition at time $t=0$ with the particle having been held at the
site $|R\rangle$ ($\sigma_z=+1$) from time $t_0=-\infty$ till
$t=0$, and with the bath in thermal equilibrium. Then the exact formal solution for $P(t)$  can be
expressed in terms of a real time double path integral over
forward $\sigma(\tau)$ and backward
$\sigma'(\tau)$ spin paths \cite{Weiss99,PhysRep98} with piecewise constant values
$\pm 1$. 
\begin{figure}[t!]
\includegraphics[width=0.48\textwidth,bb=100 580 450 730,clip=true]{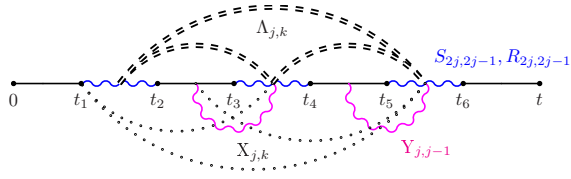}%bb=100 550 400 750,clip=true
\vspace{-1.7cm}
\caption{Generic path with $2n=6$ transitions at flip times
$t_{1},t_2,..t_{2n}$. The system is in an off-diagonal state
(blip) of the reduced density matrix 
 in the time intervals $\tau_j \equiv t_{2j}-t_{2j-1}$  and in a diagonal state (sojourn) at times
 $t_{2j+1}-t_{2j}$.
   The interactions $S_{2j,2j-1}$, $R_{2j,2j-1}$ and $Y_{j,j-1}$ (intra-dipole and blip-preceeding-sojourn interactions), Eq. \eqref{influence}, are
symbolized by the wiggled lines (blue and magenta online, respectively). The double-dashed lines denote the
inter-dipole interactions $\Lambda_{j,k}$, while the bold-dotted lines are the remaining
blip-sojourn interactions $X_{j,k}$, cf. Eq. \eqref{influencei}.%\vspace{-0.5cm}
\label{Fig.interactionsP}}
\end{figure}
Upon introducing the linear combinations
$\eta(\tau)/\xi(\tau)=[\sigma(\tau)\pm\sigma'(\tau)]/2$, one finds
\begin{equation}
P(t)=\int{\cal D}\xi{\cal D}\eta {\cal
A}[\xi,\eta]\exp{\{\Phi[\xi,\eta]\}}\;,
\end{equation}
where ${\cal A}$ is the path weight in the absence of the bath
coupling. 
A generic double path can now be visualized as a single path over the four-states of the reduced density matrix, characterized by $(\eta(\tau)=\pm 1\,, \xi(\tau)=0)$ and $(\eta(\tau)=0\,, \xi(\tau)=\pm 1)$. The time intervals spent~in a diagonal $(\xi(\tau)=0)$ and off-diagonal $(\eta(\tau)=0)$ state are dubbed ``sojourns'' and ``blips'', respectively \cite{Leggett87}.
Due to the initial condition, the path sum runs over all
paths with boundary conditions $\xi(0)=\xi (t)=0$ and $\eta(0)=1$, $\eta(t)=\pm 1$. 
% Environmental effects are in an influence
% functional inducing
%  non-local in time correlations among different path~segments:
Environmental effects are in the %influence
functional
%It reads
\vspace{-0.1cm}\begin{equation} \label{infl}
\Phi[\xi,\eta] \equiv \int_0^t \!\!dt_2 \int_0^{t_2} \!\!\!dt_1 \,\dot{\xi}(t_2)\left[ S_{2,1} \dot{\xi}(t_1)+ i R_{2,1} \dot{\eta}(t_1) \right]\!\!,\vspace{-0.1cm}
\end{equation}
with the bath correlation function $Q=S+i R$ being
\vspace{-0.11cm}\begin{equation}
Q(t)=\!\!\int_0^\infty \!\!\!\!d\omega
\frac{G(\omega)}{\omega^2}\Big[\coth\Bigl(\frac{\hbar\omega}{2k_{\rm B}T}\Bigr)(1-\cos\omega
t )+i\sin\omega t\Big]
\end{equation}
and $Q_{j,k}:=Q(t_j-t_k)$. % and $t_{2n+1}:=t$.
%
% Let us  first focus   on the evolution of   diagonal elements.
For a generic path with $2n $ transitions at times $t_j$, $j=1,2,...,2n$, one finds $\dot
\xi(\tau)=\sum_{j=1}^{2n}\xi_j\delta(\tau-t_j)$ and $\dot
\eta(\tau)=\sum_{j=0}^{2n}\eta_j\delta(\tau-t_j)$. Here is $\eta_0=1$
due to the initial preparation and $\xi_j =\pm 1$, $\eta_j=\pm 1$ for $j>0$.
Because $\xi_{2j}=-\xi_{2j-1}$,  the
influence function in \eqref{infl} becomes $\Phi^{(n)}=\Phi_{\rm intra, bps}^{(n)}+\Phi_{\rm
inter}^{(n)}$ (Fig. \ref{Fig.interactionsP}). The function $\Phi_{\rm intra, bps}^{(n)}$ describes intra-blip and blip-preceeding sojourn correlations, and reads
\begin{align}
\begin{split}
\Phi_{\rm intra, bps}^{(n)}& =  - \sum_{j=1}^n\Bigl[S_{2j,2j-1}-i
\xi_j\eta_{j-1}   X_{j,j-1}    \Bigr]\label{influence}\\
& = \Phi_{\rm intra}^{(n)}+ \Phi_{\rm bps}^{(n)}
\;,
\end{split}\\
\Phi_{\rm intra}^{(n)} &=  - \sum_{j=1}^n\Bigl[S_{2j,2j-1}-i
\xi_j\eta_{j-1}   R_{2j,2j-1}    \Bigr] \,, \label{intra}\\
\Phi_{\rm bps}^{(n)} &=  i \sum_{j=1}^n
\xi_j\eta_{j-1}   Y_{j,j-1} \;,   \label{bps}
%\end{split}
\end{align}
where we split $X_{j,j-1}= R_{2j,2j-1}+ Y_{j,j-1}$, with 
\begin{align} \label{yjj-1}
 Y_{j,j-1} = R_{2j-1,2j-2}-R_{2j,2j-2}\,.
\end{align}

Moreover, the functional $\Phi_{\rm
inter}^{(n)}$ accounts for inter-blip and blip-sojourns interactions \cite{Leggett87,Weiss99} 
\begin{gather}
\Phi_{\rm inter}^{(n)} =
-\sum_{j=2}^n\sum_{k=1}^{j-1}\xi_j\xi_k\Lambda_{j,k}+i\sum_{j=2}^n
\sum_{k=0}^{j-2}\xi_j\eta_k X_{j,k}\,. \label{influencei}
\end{gather}

\begin{figure}[t!]
\centering
\includegraphics[width=0.49\textwidth,bb=80 590 440 700,clip=true]{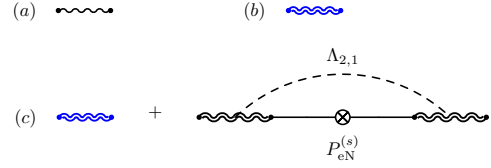}%bb=100 550 400 750,clip=true
\vspace{-0.65cm}
\caption{Irreducible kernel $K^{(s)}(t,t')$ in the
NIBA (a),  in the \textit{extended}-NIBA (blue online) (b) and in the WIBA (c). The single-dashed
%and dotted 
 lines are the {\em linearized} blip-blip %and sojourn-blip
 interactions between the first and last dipole. The
inner bubble denotes the infinite sum of \textit{extended}-NIBA diagrams yielding
the symmetric part of $P(t)$ within the \textit{extended}-NIBA, denoted $P^{(s)}_{{\rm eN}}$. %\vspace{-0.15cm}
\label{Fig.interactions}}
\end{figure}
%\end{fmffile}

The function $\Lambda_{j,k}$  contains the blip-blip interactions
between
 the flip pairs $\{j,k\}$, while the
blip-sojourn  interaction $X_{j,k}$ yields a phase factor.  To be
definite, for $k>0$,
\begin{subequations}
\begin{align}
%\begin{split}
\hspace{-0.2cm}\Lambda_{j,k}\!&=S_{2j,2k-1}+S_{2j-1,2k}-S_{2j,2k}-S_{2j-1,2k-1}, \label{lint}\\
\hspace{-0.2cm}X_{j,k}\!&=R_{2j,2k+1}+R_{2j-1,2k}-R_{2j,2k}-R_{2j-1,2k+1}\,. \label{xint}
%\end{split}
\end{align}
\end{subequations}

% For a generic path with $2n $ transitions at times $t_j$, $j=1,2,...,2n$, one finds $\dot
% \xi(\tau)=\sum_{j=1}^{2n}\xi_j\delta(\tau-t_j)$ and $\dot
% \eta(\tau)=\sum_{j=0}^{2n}\eta_j\delta(t-t_j)$. Here is $\eta_0=1$
% due to the initial preparation and $\xi_j (\eta_j) =\pm 1$.
% Because $\xi_{2j}=-\xi_{2j-1}$,  the
% influence function in \eqref{infl} becomes $\Phi^{(n)}=\Phi_{\rm intra, bps}^{(n)}+\Phi_{\rm
% inter}^{(n)}$, where $\Phi_{\rm intra, bps}^{(n)}$ describes intra-blip and blip-preceeding sojourn correlations, and $\Phi_{\rm
% inter}^{(n)}$ accounts for inter-blip interactions \cite{Leggett87,Weiss99} (Fig. 1)
% \begin{gather}
% %\begin{split}
% \Phi_{\rm intra, bps}^{(n)} =  - \sum_{j=1}^n\Bigl[S_{2j,2j-1}-i
% \xi_j\eta_{j-1}   X_{j,j-1}    \Bigr]
% \;,\label{influence}\\
% \Phi_{\rm inter}^{(n)} =
% -\sum_{j=2}^n\sum_{k=1}^{j-1}\xi_j\xi_k\Lambda_{j,k}+i\sum_{j=2}^n
% \sum_{k=0}^{j-2}\xi_j\eta_k X_{j,k} \label{influencei}\;.
% %\end{split}
% \end{gather}
% The function $\Lambda_{j,k}$  contains the blip-blip interactions
% between
%  the flip pairs $\{j,k\}$, while the
% blip-sojourn  interaction $X_{j,k}$ yields a phase factor.  To be
% definite, for $k>0$,
% \begin{equation}
% \begin{split}
% X_{j,k}\!&=R_{2j,2k+1}+R_{2j-1,2k}-R_{2j,2k}-R_{2j-1,2k+1}\;,\\
% \Lambda_{j,k}\!&=S_{2j,2k-1}+S_{2j-1,2k}-S_{2j,2k}-S_{2j-1,2k-1}
% \;.
% \end{split}
% \end{equation}
% %
 The correlations $X_{j,0}$  depend on the
 initial preparation \cite{Weiss99}.
% $\eta (\tau)$ characterizes propagation along the
% diagonal of the RDM,  while $\xi (\tau)$ records excursions in the off-diagonal
%states.
% Similar considerations hold for the RDM coherences.
 The summation over the path histories then reduces to an expansion in the number of tunneling transitions yielding  formally exact, but practically untractable, equations
for $P(t)$ of the form (\ref{GME}) \cite{EPJB99}. 

\section{Known and novel approximation schemes}

To find appropriate approximation schemes to
the TLS dynamics, let us start
 from the  familiar
 non-interacting-blip approximation  (NIBA)
 \cite{Leggett87,Weiss99}. Within the NIBA, one sets $\Phi_{\rm inter}^{(n)}=0$, namely
the inter-blip correlations  $\Lambda_{j,k}$
 and the  blip-sojourn interactions  $X_{j,k}$ ($k\ne j-1$) are
 neglected. The blip-preceeding-sojourn interactions $Y_{j,j-1}$ in Eq. \eqref{yjj-1} are neglected as well. Hence, $X_{j,j-1}$
 reduces to $X_{j,j-1}\approx R_{2j,2j-1}$.
The influence function (\ref{influence}) then 
%factorizes 
splits into individual
 influence factors depending only on the dipole length
 $\tau_j:=t_{2j}-t_{2j-1}$.
The dynamics is thus described by (\ref{GME}) with NIBA kernels   corresponding to   the one-dipole irreducible contributions
% \cite{Weiss99,PhysRep98}, 
(Fig. 2a),
\begin{equation} \label{kernNIBA}
\begin{split}
K^{s}_{{\rm
N}}(t,t')&=\Delta^2 {\cal C}(t-t')\cos[\zeta(t,t')]\;,\\
K^a_{{\rm N}}(t,t')&=\Delta^2 {\cal
S}(t-t')\sin[\zeta(t,t')]\;,
\end{split}
\end{equation}
with $\zeta(t,t')=\int^t_{t'}dt''\varepsilon (t'')$,  intra-blip
contributions
 ${\cal C}(t)=e^{-S(t)}\cos[R(t)]$, and ${\cal S}(t)=e^{-S(t)}\sin[R(t)]$. Here, $W_{\rm N}=0$.
% Likewise, $K^{s}_{x,{\rm N}}(t,t')=\Delta\, {\cal S}(t-t')\cos[\zeta(t,t')]$, $K^{a}_{x,{\rm N}}(t,t')=\Delta \,{\cal C}(t-t')\sin[\zeta(t,t')]$.
 The kernels are of lowest order in the tunneling matrix $\Delta$
 but are {\em non}-perturbative in $\delta_s$.
Due to the simplicity of the kernels \eqref{kernNIBA}, the NIBA has been a very popular approximation so far. 
% For
%time-independent bias the equations (\ref{GME}) are easily solved
%by Laplace transformation.
 For sub-Ohmic damping, $s<1$, NIBA is expected to be a valid approximation for all temperatures with the TLS  exhibiting incoherent dynamics even for very small
coupling $\delta_s$. For Ohmic and super-Ohmic damping, NIBA is
expected to be a good approximation only at high enough temperature
 and/or strong damping \cite{Weiss99}. 
%\footnote{Ohmic spectral densities reach for large temperatures faster the asymptotic
% behavior $S(t)\propto t$ at long times, implying that the
% correlations in $\Lambda_{j,k}$ cancel out exactly.}.  
However, its limit of validity are \textit{not clearly} defined. The NIBA is
known to fail at low temperatures and weak coupling for
an asymmetric TLS for Ohmic and super-Ohmic damping, because the
dipole-dipole correlations $\Lambda_{j,k}$ contribute already to
terms which depend linearly on the spectral density $G(\omega)$.
For example, in the case of a TLS with static asymmetry $\varepsilon(t)=\varepsilon_0$, NIBA predicts the unphysical asymptotic limit
\begin{equation}
P_{{\rm N}}^{\infty}=-\tanh
(\frac{\beta\hbar\varepsilon_0}{2}) \,,
\end{equation}
implying
 localization of the TLS ($P_{{\rm N}}^\infty=-1$) at
zero temperature even for infinitesimal asymmetries. 
In order to overcome the NIBA shortcomings, a weak-coupling approximation (WCA) has been proposed in \cite{Weiss99,EPJB99,Görlich89} with WCA kernels being linear in $\delta_s$ and~nonperturbative~in~$\Delta$. Within the WCA, the TLS dynamics
shows damped coherent oscillations with a renormalized energy
splitting $\hbar\Omega$, with $\Omega^2=\Delta^2_{\rm eff}[1-2 {\rm Re}
\,u(iE/\hbar)] +\varepsilon_0^2$, towards the equilibrium value
\begin{equation}
P_{{\rm WCA}}^{\infty}=-\frac{\hbar\varepsilon_0}{E}\tanh
(\frac{\beta E}{2})\,. 
\end{equation}
Here, the frequency shift is related to the frequency
integral $u(z)=\frac{1}{2}\int_0^\infty d\omega
\frac{G(\omega)}{\omega^2+z^2}[\coth(\hbar\beta\omega/2)-1]$. Moreover, $E=\hbar\sqrt{\Delta_{\rm eff}^2+\varepsilon_0^2}$ and 
the effective bath-renormalized tunneling coupling $\Delta_{\rm eff}$ for the cases $s\ge 1$ reads
%In the sub-Ohmic case incoherent dynamics is
%expected even for very small $\delta_s$.
\cite{Weiss99}
\vspace{-0.1cm}\begin{eqnarray}
\Delta_{\rm eff}&=&\Delta[\Gamma(1-2\alpha)\cos(\pi
\alpha)]^{\frac{1}{2(1-\alpha)}}\left(\Delta/\omega_c\right)^{\frac{\alpha}{1-\alpha}},
\quad s=1\,, \nonumber\\
\Delta_{\rm eff}&=&\Delta\exp[\delta_s\Gamma
(s-1)(\omega_{c}/\omega_{\rm{ph}})^{s-1}],\qquad s>1 \;,\vspace{-0.1cm}\end{eqnarray}
with $\Gamma(z)$ the Gamma function. 
Finally, the relaxation $\Gamma_r$ and dephasing $\Gamma_\phi$
rates are given  by the perturbative (in $\delta_s$) expressions $\Gamma_r=  (\pi \hbar^2\Delta^2_{\rm eff}/2 E^2)A(E/\hbar)$ and $\Gamma_\phi=  \Gamma_r/2 +(\pi\hbar^2\varepsilon_0^2/2E^2)A(0)$, 
where the spectral function
$A(\omega)=G(\omega)\coth(\hbar\omega/2k_{\rm B}T)$ is related to
emission and absorption of a single phonon. 

To smoothly bridge between the high and low $T$ limits, let us now start to discuss a more refined approximation, which we call \textit{extended}-NIBA (Fig. 2b). As in NIBA, $\Phi_{\rm inter}$ is neglected,  while the approximation on the blip-preceeding sojourn interactions $X_{j,j-1}$ is improved, considering also $Y_{j,j-1}$ in an effective way. Specifically, expanding $Y_{j,j-1}=-R_{2j,2j-2}+R_{2j-1,2j-2}$ up to first order in the blip lengths $\tau_j$, we set $X_{j,j-1} \approx R(\tau_j)-\tau_j \dot{R}_{2j-1,2j-2}$. As a result, the \textit{extended}-NIBA kernels $K^{s/a}_{{\rm eN}}(t,t')$ have the same form as \eqref{kernNIBA} with ${\cal C} (t) \to {\cal C'} (t)
% $K^{s}_{z,{\rm eN}}(t,t')=\Delta^2 {\cal C'}(t-t')\cos[\zeta(t,t')]$ and $K^a_{z,{\rm eN}}(t,t')\equiv K^a_{z,{\rm N}}(t,t')$, where
=e^{-S(t)}\cos[\tilde{R}(t)]$. Here is $\tilde{R}(t) \equiv R(t)-t\,\dot{R}(t)$. Moreover, $W_{\rm eN}=K^{s}_{{\rm eN}}-K^{s}_{{\rm N}}$.
A comparison between NIBA and \textit{extended}-NIBA, as well as with other approximation schemes discussed below, is shown in Figs. 3a - 3d. In Figs. 3b - 3d we also show results obtained with the numerical ab-initio path-integral approach QUAPI~\cite{Makri95}. % are shown in Figs. 3c, 3d. 
The short-time dynamics is always well approximated by the NIBA (\textit{extended}-NIBA). At long times, however, correlations neglected in the NIBA become relevant. In particular, already at intermediate temperatures and damping (Figs. 3c, 3d) the \textit{extended}-NIBA correctly reproduces the QUAPI results while NIBA fails. At low $T$ and small damping, both NIBA and \textit{extended}-NIBA fail to reproduce the correct long time dynamics (Fig. 3a), as predicted e.g. from the~WCA.

\begin{figure*}[th!]
\begin{center}
\includegraphics[width=1.\textwidth, bb=60 60 727 534,clip=true]{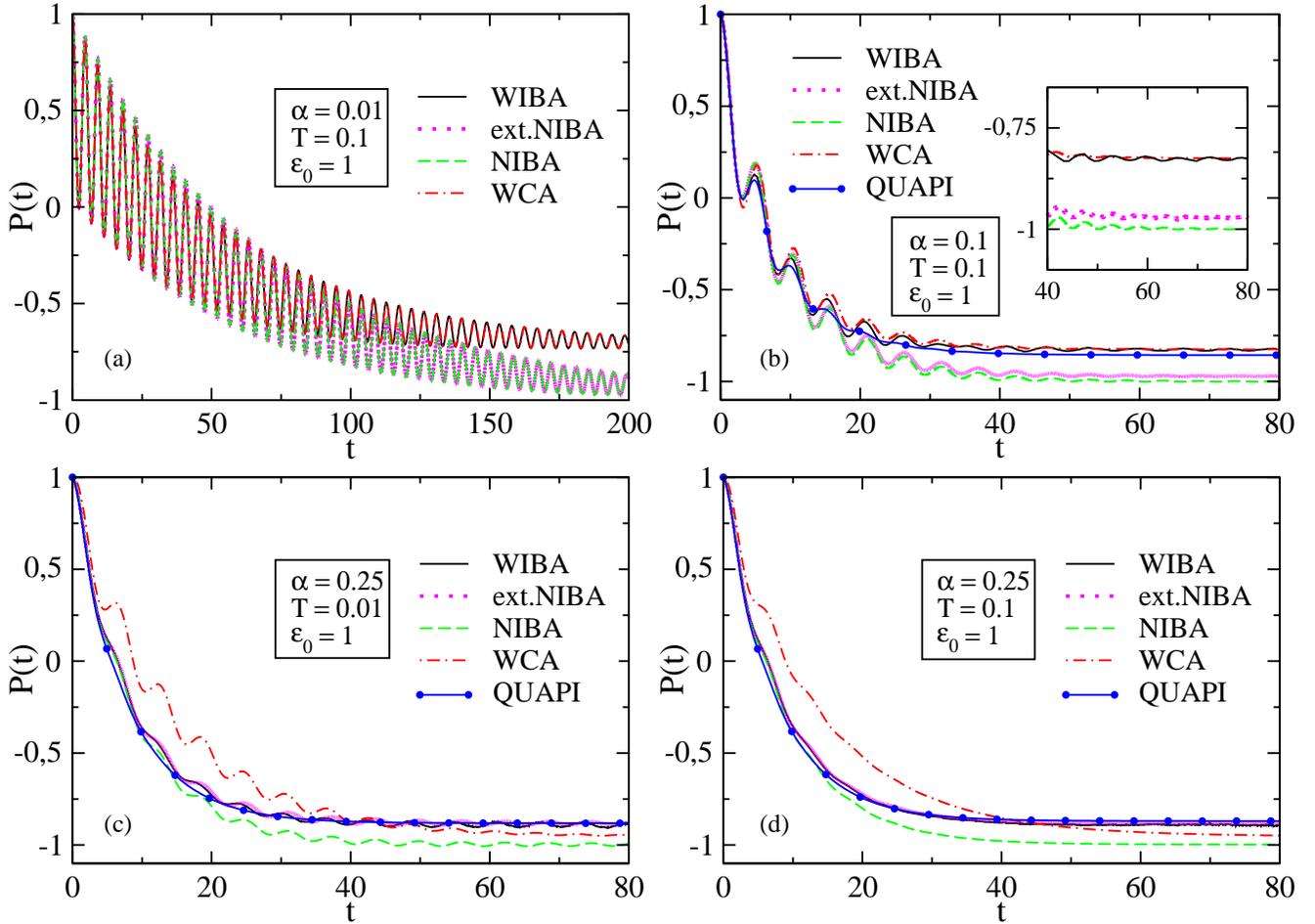}%65 75 727 526, trim=5mm 0 0 0, clip=true
%\epsffile{fig2.eps}
%\vspace{-0.3cm}
\caption{Time evolution of the expectation value $P(t)$ at
low/moderate temperatures $k_{\rm B}T \lesssim E$ for several
values of the Ohmic coupling parameter $\alpha$. Full lines
depict the WIBA, dashed lines the NIBA, dotted lines the \textit{extended}-NIBA, the dot-dashed ones are
results for the weak-coupling approach (WCA) while the lines with
bullets are the ab-initio QUAPI predictions. 
% In Fig. 2(b), $\sigma_z^{1/2}$ is the analytical solution of Eq. \eqref{1/2}. 
All quantities are
expressed in units of $\Delta$. 
At low damping and temperatures, Fig.\ 3a, the TLS exhibits damped coherent oscillations towards the asymptotic value 
$P_{\rm WCA}^\infty$. As the damping is increased, the oscillations are more strongly damped, see Figs.\ 3b, 3c, 3d.
In the chosen regime of parameters, the equilibrium value $P^\infty$ is neither well described by $P_{\rm N}^\infty$ nor by 
$P_{\rm WCA}^\infty$.
\vspace{-0.2cm}
\label{Fig.Ohmic1}}
\end{center}
\end{figure*}

To bridge between the moderate damping situation well
 described by the \textit{extended}-NIBA  and the extremely underdamped case we observe
that, for  spectral densities of the form (\ref{deltas}),   the
blip-blip interaction terms $\Lambda_{j,k}$ as well as the
blip-sojourn terms $X_{j,k}$ ($k\neq j-1$) are intrinsically small compared to
unity. %The correlations $X_{j,j-1}$ are treated in the spirit of the \textit{extended}-NIBA.
Hence, we propose a novel approximation scheme, which we
call the weakly-interacting blip approximation (WIBA).
Within the WIBA, the full $\Phi_{\rm intra,bps}^{(n)}$ is retained as in the \textit{extended}-NIBA and one expands the influence functional $\exp{\{\Phi_{\rm inter}^{(n)}\}}$ up to \textit{linear order} in the blip-blip and blip-preceeding sojourns interactions $\Lambda_{j,k}$ and $X_{j,k}$.
In other terms, 
\begin{equation} \label{wexp}
 \exp{\{\Phi^{(n)}\}}\approx \exp{\{\Phi_{\rm intra,bps}^{(n)}\}} \left(1+\Phi_{\rm inter}^{(n)}\right).
\end{equation}
%
% , where the
% intra-blip
%  correlations and blip-preceeding sojourn interactions are treated as in the \textit{extended}-NIBA, while the
% inter-blip and the other blip-sojourn interaction effects are considered up
% to {\em linear order}. 
\begin{figure*}[th!]
\begin{center}
\includegraphics[width=1.\textwidth,bb=60 294 727 534,clip=true]{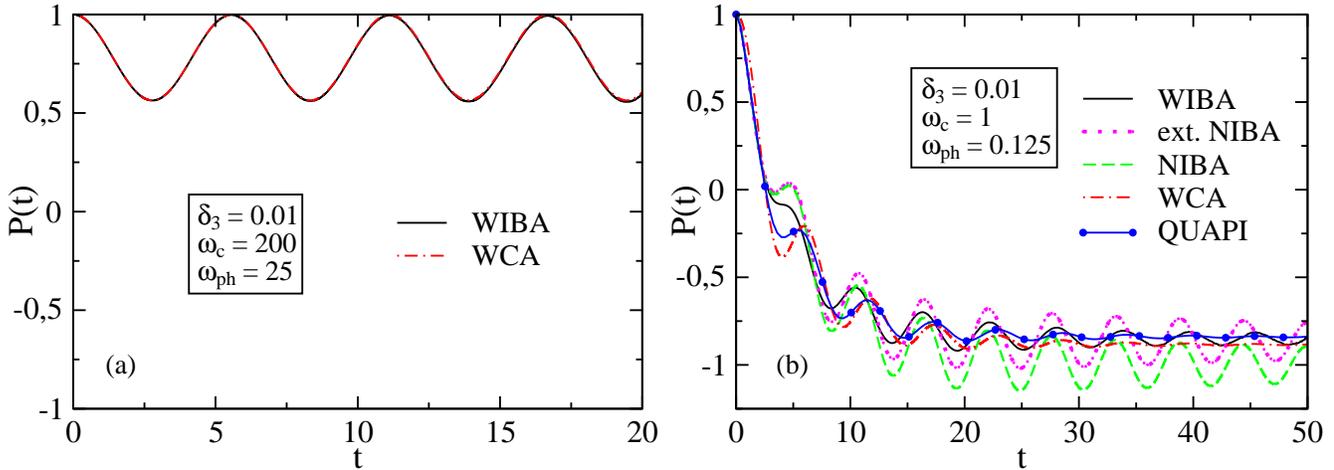}%65 75 708 526, height=6cm,bb=65 55 708 570,58 64 708 526
%\vspace{-0.3cm}
\caption{Time evolution of $P(t)$ for super-Ohmic damping
(coupling parameter $\delta_3$). Here the ratio $\omega_c/\omega_{\rm ph}=8$ is kept constant in both panels and we set $T=0.1,\:\varepsilon_0=1$ (in units of $\Delta$). Full
lines are the WIBA predictions, %, dashed lines the NIBA expectation,
dot-dashed lines are results of the
WCA, dashed lines represent the NIBA predictions, dotted lines denote the \textit{extended}-NIBA dynamics and finally bulleted lines are results of QUAPI. \vspace{-0.2cm}
\label{Fig.Ohmic3}
}
\end{center}
\end{figure*}
Within an expansion in the number of
tunneling transitions, the
 lowest order self-energy corresponds to the \textit{extended}-NIBA, while
  higher order terms describe a set of blips in which the
 first and last blip are interacting, Fig. 2c. Summing up the higher
 contributions,  the WIBA
kernels, {\em  neither perturbative in $\Delta$ nor in
$\delta_s$}, read
\begin{align} \label{WIBAs}
\lefteqn{
K^s_{{\rm W}}(t_4,t_1)= K^s_{{\rm eN}}(t_4,t_1)}
\nonumber
\\  &-  \Delta^4
\int_{t_1}^{t_4}dt_3\int_{t_1}^{t_3}dt_2 {\cal C'}(t_4-t_3)
\sin[\zeta(t_4,t_3)] P_{{\rm eN}}^s(t_3-t_2) \nonumber
\\&\times  \Lambda_{2,1}{\cal C'}(t_2-t_1)\,\sin[\zeta(t_2,t_1)]\;,%\vspace{-0.5cm}
%+X_{2,0}{\cal S}(t_2-t_1)
\end{align}
\begin{align}
 \lefteqn{K^a_{{\rm W}}(t_4,t_1)= K^a_{{\rm eN}}(t_4,t_1)}
 \nonumber\\ &- \Delta^4
\int_{t_1}^{t_4}dt_3\int_{t_1}^{t_3}dt_2  {\cal C'}(t_4-t_3)
\sin[\zeta(t_4,t_3)] P^s_{{\rm eN}}(t_3-t_2) \nonumber\\
 &\times [-\Lambda_{2,1}{\cal S}(t_2-t_1)+X_{2,0}{\cal
C}(t_2-t_1)]\cos[\zeta(t_2,t_1)] \;. \label{WIBAa}
\end{align}
\vspace{-0.1cm}
Moreover,
\begin{align}
 \lefteqn{W_{\rm W}(t_4,t_1)= W_{\rm eN}(t_4,t_1)}
 \nonumber\\ &- \Delta^4
\int_{t_1}^{t_4}dt_3\int_{t_1}^{t_3}dt_2  {\cal C'}(t_4-t_3)
\sin[\zeta(t_4,t_3)] P^s_{{\rm eN}}(t_3-t_2) \nonumber\\
 &\times [\Lambda_{2,1} \delta{\cal C}(t_2-t_1)-X_{2,0}{\cal
S}(t_2-t_1)]\sin[\zeta(t_2,t_1)] \;, \label{WIBAw}
\end{align}
where $ \delta{\cal C}\equiv {\cal C'}-{\cal C}$. 
%
% Similar expressions are obtained for the kernels $K^{s/a}_{x,{\rm W}}$ and will be given elsewhere. 
Moreover, $P^s_{{\rm
eN}}(t)$ is the symmetric part (in $\varepsilon_0$) of
$P_{{\rm eN}}(t)$   {\em within} the \textit{extended}-NIBA. Thus, at high
temperatures, where the blip-blip interactions are negligible,
the
 WIBA kernels reduce to the \textit{extended}-NIBA ones. By expanding the
 WIBA kernels to first order in $\delta_s$ and approximating $X_{j,j-1}$ to $R_{2j-2j-1}$, the weak damping
  kernels  in \cite{Weiss99,EPJB99} are recovered.

\section{Ohmic damping}

As a benchmark for the WIBA, we consider the evolution of the population difference $P(t)$ for the important case of Ohmic damping.  %static asymmetry $\varepsilon(t) =\varepsilon_0$,
%
% A comparison between
% the WIBA, the NIBA and the analytical expression WCA for 
In Fig. 3, the Ohmic %and super-Ohmic (with $s=3$)
case ($\omega_c=50\Delta$ and $\varepsilon(t)=\varepsilon_0=\Delta$) is shown. An excellent agreement is found for
weak damping and temperatures (Fig. 3a) between WIBA and WCA, whereas the \textit{extended}-NIBA matches the NIBA %, since the coupling is so small that the extra-correlations included in the extension do not really make any difference.
and predicts  the wrong asymptotic limit
$P_{{\rm N}}^{\infty}$. %, because in this regime of parameters $W$ has a vanishing contribution in the \textit{extended}-NIBA. 
As the coupling is increased (or by
raising the temperature),
% two-phonon and higher order processes
%become relevant and
 the WCA is expected to fail. However, as for the NIBA, the limits of validity of the WCA are not clearly defined. Indeed, Figs. 3b to 3d
show an intermediate parameter regime where both approximations 
fail, since dipole-dipole interactions as well as two-phonon
processes are relevant. Comparison with results from QUAPI shows that
the short time dynamics is well approximated by the NIBA (WIBA).
At intermediate and long times, the WIBA reasonably well approaches QUAPI and its asymptotic value. 
From a comparison with QUAPI, we notice
that the higher order dipole correlations neglected in the WIBA
yield a larger dephasing rate than predicted from WIBA. In particular, QUAPI predicts 
a complete suppression of the coherent oscillations already
at $\alpha=0.25, T=0.01$.
%
% The \textit{extended}-NIBA also matches perfectly the WIBA behavior, showing that for $\alpha=0.25$ the extra-correlations included in this model are sufficient to reproduce the WIBA results.
%
An interesting case is shown in Fig. 3b, with a small-to-intermediate value of the coupling strength ($\alpha=0.1$), where the \textit{extended}-NIBA slightly moves from the NIBA towards the WIBA predictions, reaching an intermediate asymptotic value (see inset).

\section{Super-Ohmic damping}

Let us now consider the predictions of the WIBA
in the super-Ohmic case ($s=3$). 
Since $S(t \sim \Delta^{-1})$ differs only little from its asymptotic value $S(t \to \infty)$, the interblip interactions are weak and the WCA is expected to be a good approximation in a wide regime of parameters. This also implies that $S(t)$ is \textit{not} effective in suppressing long-blip lenghts and the NIBA might not be justified. Indeed, 
for small $\delta_s$ and large $\omega_c$ (see Fig.\ 4a), no differences among WCA
and WIBA occur. 
%Notice the deviation from the numerical result of QUAPI when, keeping the ratio $\omega_c/\omega_{\rm ph}$ constant, 
Similarly to the Ohmic case, we show in Fig.\ 4b the parameter regime where the WCA and the NIBA are expected to fail. With respect to Fig.\ 4a, we keep here the same ratio $\omega_c/\omega_{\rm ph}$ constant, being now $\omega_c \sim \Delta$, i.e. the bath becoming ``slow''. 
This case is the most difficult one, since the bath is very coherent and memory effects are to be taken into account, which requires to perform a very good description of the full bath dynamics.
One sees that the NIBA completely fails to reproduce the dynamics, even reaching unphysical values. The \textit{extended}-NIBA works better, approaching closer the QUAPI predictions. Nevertheless, too few correlations are taken into account, and it oscillates still too much with respect to the numerical plot of QUAPI. The WIBA shows discrepancies from the QUAPI as well, being still ``too'' coherent, 
even though its predictions are  more accurate than the \textit{extended}-NIBA.
 The WCA, despite better than WIBA in this regime, also lies apart from the numerical prediction of QUAPI. In this range of parameters, the multiphonon processes become relevant and the perturbative weak-coupling approximation begins to fail.
This agrees with ab-initio simulations for charge qubits interacting with piezoelectric phonons \cite{Vorojtsov05}.
Hence, further analysis of the complicated super-Ohmic case is to be done, in order to better understand the different dynamical situations which take place by varying the coupling strength $\delta_s$, the cutoff frequency $\omega_c$ and the phonon frequency $\omega_{\rm ph}$.

\section{Conclusions}

We have discussed a generalized master equation for the population difference $P(t)$ of a spin-boson system
    in the whole regime of temperatures and couplings. 
%for a wide class of spectral densities (Eq. \eqref{deltas}). %(also for time-dependent driving).
%
This equation can be solved using standard iteration schemes \cite{Hartmann00,Annals}. 
%     It nicely reproduces known results in various
%     complementary regimes where it yields the correct asymptotic value. The GME can be used to investigate the spin-boson dynamics also in regimes where, to
%     date, only numerical ab-initio approaches are available.
For Ohmic damping the WIBA is able to reproduce known results in various complementary regimes, yielding a good, though not perfect agreement, with \textit{ab-initio} QUAPI calculation in the regime of intermediate temperatures and damping. 
Hence, it  overcomes the limits of validity of the perturbative approaches (NIBA, WCA) which, up to date, was possible only with numerical ab-initio models. 
For super-Ohmic damping the WIBA works well for large cut-off frequencies and low-to-moderate temperatures. However, disagreement with QUAPI is found in the case of a ``slow'' bath.

We mention some general contexts for the need of a bridging approach: i) The common experimental situation where bath temperature or TLS asymmetry are varied over a wide range (WCA and NIBA are unreliable at high temperatures and intermediate bias, respectively). ii) Several TLS's interacting with a common heat bath, as e.g. in glasses at low-temperatures \cite{glass, glass2}.
Due to the wide distribution of tunneling parameters and asymmetries, neither the WCA nor the NIBA can describe the dynamics of the whole ensemble consistently. 

We must, however, notice that in the case of ``slow'' environments with cut-off frequency of the order of the tunneling frequency, our model needs some improvements, since neither the WIBA nor other analytical approximation schemes are able to reproduce the correct onset of decoherence which in fact takes place.
 This situation occurs e.g. in  non-adiabatic electron transfer \cite{Garg85} or for charge qubits interacting with piezo-electric phonons \cite{Vorojtsov05}.
%
% Therefore, the WIBA equations become of great relevance in the very common situation of TLSs which can undergo a transition from coherent to incoherent tunneling by simply varying the temperature or the bias. Moreover, they  can also be applied to systems like glasses (metallic or insulating) \cite{glass}, which possess a broad distribution of asymmetries and tunneling parameters, and to ``slow'' baths (i.e. with cut-off frequencies $\omega_c \sim \Delta$).
% Finally, the WIBA can also become essential for ``slow'' baths (i.e. with cut-off frequencies $\omega_c \sim \Delta$). In this situation, which may occur e.g. in double quantum dots subject to piezoelectric phonons, ab-initio simulations \cite{Vorojtsov05} showed that a perturbative approach in the coupling constant is going to fail.

%
%\vspace{0.15cm}
%\centerline{***}
%\vspace{-0.15cm}

\acknowledgments
\section{Acknowledgments}
 Discussions with L. Hartmann, D. Bercioux, M. Storcz, A. Donarini and support  under the DFG programs GKR638 and SFB631 are acknowledged.
%\vspace{-0.55cm}
%

\end{document}